\documentclass[twocolumn,letter]{jpsj2}
\def\runtitle{Quantum Electron Transport in Disordered Wires
with Symplectic Symmetry}
\def\runauthor{Yositake {\sc Takane}}
\title{%
Quantum Electron Transport in Disordered Wires with Symplectic Symmetry
}

\author{%
Yositake {\sc Takane}
}

\inst{%
Department of Quantum Matter, Graduate School of Advanced Sciences of Matter,
Hiroshima University, Higashi-Hiroshima 739-8530, Japan
}

\abst{%

The conductance of disordered wires with symplectic symmetry
is studied by the supersymmetric field theory.
Special attention is focused on the case
where the number of conducting channels is odd.
Such a situation can be realized in metallic carbon nanotubes.
The average dimensionless conductance $\langle g \rangle$
is obtained using Zirnbauer's super-Fourier analysis.
It is shown that with increasing wire length,
$\langle g \rangle \to 1$ in the odd-channel case,
while $\langle g \rangle \to 0$ in the ordinary even-channel case.
It should be emphasized that the so-called Zirnbauer's zero mode,
which has been believed to be unphysical, is essential
for describing the anomalous behavior in the odd-channel case.
}

\kword{%
supersymmetric field theory, nonlinear $\sigma$ model, DMPK equation,
carbon nanotube
}

\begin{document}
\sloppy
\maketitle

Phase-coherent electron transport in quasi-one-dimensional disordered
wires is characterized by three universality classes.~\cite{beenakker}
The universality classes describe transport properties
which are independent of the microscopic details of disordered wires.
The orthogonal class consists of systems having time-reversal symmetry
without spin-orbit scattering, while the unitary class is
characterized by the absence of time-reversal symmetry.
The systems having time-reversal symmetry with strong spin-orbit scattering
belong to the symplectic class.
The conductance as well as its moment averaged over an ensemble
crucially depends on which universality class it belongs to.
For example, weak-localization (weak-antilocalization) effect arises
in the orthogonal (symplectic) class,
while such effects disappear in the unitary class.
We can systematically analyze such differences
by using two different nonperturbative approaches.

The first is the scaling theory of Dorokhov,~\cite{dorokhov}
and Mello, Pereyra and Kumar.~\cite{mello}
Let us consider an ensemble of disordered wires of length $L$
with $N$ channels per spin.
The conductance of a sample is expressed as $G = (2e^{2}/h) g$ with
$g = \sum_{a=1}^{N} T_{a}$,
where $\{T_{a}\}$ denote the transmission eigenvalues
and the factor two in $G$ corresponds to the electron spin.
It is convenient to introduce $\lambda_{a} \equiv (1-T_{a})/T_{a}$.
The statistical properties of $g$ are determined by the probability
distribution $P(\lambda_{1},\lambda_{2},\cdots,\lambda_{N},L)$.
The scaling theory is based on an evolution equation
for $P$ with increasing $L$,
which is called the Dorokov-Mello-Pereyra-Kumar (DMPK) equation.
The second is the supersymmetric field (SSF) theory of Efetov.~\cite{efetov}
In this approach, we express the conductance (or its moment) in terms of
electron Green functions and rewrite its ensemble average in an integral form
over matrices $Q$ containing commuting and anticommuting variables.
Consequently, the problem is reduced to a field theory
with the one-dimensional nonlinear $\sigma$ model.
This approach enables us to calculate the average conductance and its variance
by means of the super-Fourier analysis by Zirnbauer.~\cite{zirnbauer,mirlin}

The equivalence of the two approaches had been questioned
because, for the symplectic class, the Zirnbauer's result indicates
$\langle g \rangle \to 1/2$ as $L \to \infty$
while the DMPK equation gives $\langle g \rangle \to 0$.~\cite{comment}
It had been recognized that the nonzero limiting value of $1/2$
is caused by the so-called Zirnbauer's zero mode,
which appears in the super-Fourier analysis.
This apparent controversy is resolved by Brouwer and Frahm.~\cite{brouwer}
They pointed out that the zero mode must be excluded
since its parity is incompatible with an appropriate initial condition.
They showed that the super-Fourier analysis gives us the result
which agrees with the DMPK approach if we exclude unphysical modes,
including the zero mode,
and correctly take Kramers degeneracy into account.

Recently, Ando and Suzuura~\cite{ando} showed that metallic carbon nanotubes
(CNs) have an anomalous transport property, which cannot be
explained within the ordinary universality classes.
They studied CNs of length $L$ when the potential range of scatterers
is larger than the lattice constant,
in the case where the Fermi level lies in several conducting channels.
Note that the number of conducting channels is odd in CNs.
They proved that one perfectly conducting channel is present
even when $L \to \infty$.
The present author~\cite{takane1} derived the DMPK equation for this system
based on the peculiar symmetry of the scattering matrix,~\cite{ando}
and found that the system belongs to the symplectic class
with an odd number of channels.
A unit cell of a CN contains two sublattices,
and this internal degree of freedom plays a role of a pseudospin.
When the potential range of scatterers is larger than the lattice constant,
there appears an additional symmetry,
which drives the system into the symplectic class.~\cite{suzuura}
Thus, the symplectic symmetry in CNs has no connection with
true spin as well as spin-orbit scattering.
Previously, the transport property in the symplectic class
has been considered only in the even-channel case.~\cite{beenakker}
However, the odd-channel case~\cite{ando,takane1,takane2,takane3} is
very different from the ordinary even-channel case.

In this letter, we consider the average conductance of disordered wires
with symplectic symmetry by means of the SSF theory.
Our attention is focused on the case where the number $N$
of conducting channels is odd.
We calculate the average conductance in the long-$L$ limit
by using the super-Fourier analysis.
In this approach,
we expand the so-called heat kernel in terms of the eigenfunctions
of the Laplacian in the space in which the $\sigma$ model is defined.
We point out that appropriate eigenmodes for the expansion
differ between the even- and odd-channel cases.
Consequently, unphysical modes in the even-channel case
turn out to be appropriate in the odd-channel case.
It is shown that the Zirnbauer's zero mode, which must be excluded
in the even-channel case, becomes essential in describing
the anomalous behavior in the odd-channel case.
If we include the appropriate modes and correctly take Kramers degeneracy
into account, we obtain $\langle g \rangle \sim 1+2 {\rm e}^{-4L/\xi}$
in the odd-channel case when $L \gg \xi$,
where $\xi \equiv 2(N-1)l$ with the mean free path $l$.
The result is in agreement with that obtained from
the DMPK equation.~\cite{takane1}
We hereafter treat only the symplectic class,
and do not consider true spin.
Our argument closely follows that by Brouwer and Frahm.~\cite{brouwer}
We first introduce a generating function $F$ for conductance,
and show that the averaged $F$ obtained by the SSF approach becomes
equivalent to that by the DMPK approach.
By inspecting this argument, we show which eigenmodes are appropriate
in expanding the heat kernel and then obtain the average conductance.

If the number of the conducting channels is even (i.e., $N = 2 m$),
the dimensionless conductance $g$ is expressed as
$g = 2 \sum_{a=1}^{m}(1+\lambda_{a})^{-1}$,
where the factor two represents Kramers degeneracy.
Extending Rejaei's idea,~\cite{rejaei}
Brouwer and Frahm~\cite{brouwer} introduced the generating function,
\begin{equation}
    \label{eq:gener-f}
   F_{0}(\hat{\theta}, \hat{\lambda})
    = \prod_{a=1}^{m} f(\hat{\theta}, \lambda_{a})
\end{equation}
with
\begin{equation}
   f(\hat{\theta}, \lambda)
    = \frac{\left( 1 + 2\lambda + \cos(\theta_{3}+\theta_{4}) \right)
            \left( 1 + 2\lambda + \cos(\theta_{3}-\theta_{4}) \right)}
           {\left( 1 + 2\lambda + \cosh\theta_{1} \right)^{2}} .
\end{equation}
Here, $\hat{\theta} = (\theta_{1},\theta_{3},\theta_{4})$
and $\hat{\lambda}=(\lambda_{1},\lambda_{2},\cdots,\lambda_{m})$.
Once the ensemble average of $F_{0}(\hat{\theta}, \hat{\lambda})$
is presented, we can calculate the average conductance, its variance and
the density of transmission eigenvalues.~\cite{brouwer}
When the number of channels is odd (i.e., $N = 2m +1$),
only one eigenvalue is not degenerate.
We assign the $N$th eigenvalue for it without loss of generality and obtain
$g = (1+\lambda_{N})^{-1}+ 2 \sum_{a=1}^{m}(1+\lambda_{a})^{-1}$.
In this case, we consider the following generating function
\begin{equation}
      \label{eq:g-func1}
   F(\hat{\theta}, \hat{\lambda})
    = f(\hat{\theta}, \lambda_{N})^{\frac{1}{2}}
      \times F_{0}(\hat{\theta}, \hat{\lambda}) ,
\end{equation}
instead of eq.~(\ref{eq:gener-f}).

Before presenting the SSF treatment, we derive the evolution equation for
the averaged generating function using the DMPK equation.~\cite{takane1}
The ensemble average of $F$ is given by
\begin{align}
     \label{eq:av-dmpk}
  \langle F(\hat{\theta}, \hat{\lambda}) \rangle_{\rm DMPK}
  & = \int_{0}^{\infty} \prod_{a=1}^{m}{\rm d}\lambda_{a}
      \int_{0}^{\infty} {\rm d}\lambda_{N}
           \nonumber \\
  &   \hspace{10mm} \times
      \delta(\lambda_{N})
      F(\hat{\theta}, \hat{\lambda})P(\hat{\lambda},L) ,
\end{align}
where $P(\hat{\lambda},L)=P(\lambda_{1},\lambda_{2},\cdots,\lambda_{m},L)$.
The probability distribution obeys the DMPK equation,~\cite{takane1}
\begin{align}
     \label{eq:dmpk}
 &  l (N-1) \frac{\partial}{\partial L} P(\hat{\lambda},L)
           \nonumber \\
 & \hspace{5mm}
      = \sum_{a=1}^{m}
        \frac{\partial}{\partial \lambda_{a}}
        \lambda_{a}(1+\lambda_{a})J(\hat{\lambda})
        \frac{\partial}{\partial \lambda_{a}}
        J^{-1}(\hat{\lambda}) P(\hat{\lambda}, L)
\end{align}
with the mean free path $l$ and 
\begin{equation}
     \label{eq:jacob1}
  J(\hat{\lambda}) = \prod_{c=1}^{m} \lambda_{c}^{2} \times
        \prod_{a=2}^{m} \prod_{b=1}^{a-1} |\lambda_{a}-\lambda_{b}|^{4} .
\end{equation}
From eq.~(\ref{eq:dmpk}), we obtain the evolution equation,
\begin{equation}
     \label{eq:F-dmpk}
  l(N-1) \frac{\partial}{\partial L}
  \langle F(\hat{\theta}, \hat{\lambda}) \rangle_{\rm DMPK}
    = \langle D_{\hat{\lambda}} F(\hat{\theta}, \hat{\lambda})
      \rangle_{\rm DMPK} ,
\end{equation}
where
\begin{equation}
    D_{\hat{\lambda}}
  = J^{-1}(\hat{\lambda}) \sum_{a=1}^{m}
    \frac{\partial}{\partial \lambda_{a}} \lambda_{a}(1+\lambda_{a})
    J(\hat{\lambda}) \frac{\partial}{\partial \lambda_{a}} .
\end{equation}
In the limit of $L \to 0$, all the channels become perfectly conducting
(i.e., $\lambda_{a} = 0$ for any $a$).
This ballistic initial condition results in
\begin{equation}
      \label{eq:init-dmpk}
   \lim_{L \to 0} \langle F(\hat{\theta}, \hat{\lambda}) \rangle_{\rm DMPK}
  = \left( \frac{\cos \theta_{3}+\cos \theta_{4}}
                {1+\cosh \theta_{1}} \right)^{2m+1} .
\end{equation}
In the even-channel case, the exponent $2m+1$ is replaced by $2m$.

We next derive the evolution equation by using the SSF theory
based on the model by Iida, Weidenm\"uller and Zuk.~\cite{iida,altland}
The derivation has been presented by Brouwer and Frahm,~\cite{brouwer}
so we only summarize it briefly.
We assume that a quasi-one-dimensional wire consists of $K$ small grains
in series.
Each grain has $2M$ electron states including Kramers degeneracy.
The wire is coupled to two ideal leads.
We assume that the left (right) lead contains $N_{1}$ ($N_{2}$)
conducting channels.
The Hamiltonian $H$ of our system is described by
the matrix elements $H_{\mu s_{\mu} \nu s_{\nu}}^{i j}$,
where $i,j$ denote the grains ($1 \le i,j \le K$),
$\mu,\nu$ denote the electron states ($1 \le \mu,\nu \le M$)
and $s_{\mu},s_{\nu}$ denote the pseudospins.
The elements $H_{\mu s_{\mu} \nu s_{\nu}}^{i i}$,
which describe the electron states in the $i$th grain,
are independent Gaussian random variables with zero mean values and
\begin{align}
  \overline{ H_{\mu s_{\mu} \nu s_{\nu}}^{ii}
             H_{\mu' s_{\mu'} \nu' s_{\nu'}}^{ii} }
  & = \frac{v_{1}^{2}}{2M}
      \Big(
         \delta_{\mu \nu'}\delta_{\nu \mu'}
         \delta_{s_{\mu} s_{\nu'}}\delta_{s_{\nu} s_{\mu'}}
                  \nonumber \\
  & \hspace{-10mm}
       - \delta_{\mu \mu'} \delta_{\nu \nu'}
         ({\rm i}\sigma_{y})_{s_{\mu'}, s_{\mu}}
         ({\rm i}\sigma_{y})_{s_{\nu}, s_{\nu'}}
      \Big) ,
\end{align}
where $\sigma_{y}$ is the Pauli matrix.
The coupling between the adjacent grains are described by
$H_{\mu s_{\mu} \nu s_{\nu}}^{i j}$ ($j=i \pm 1$),
which are another set of Gaussian random variables with zero mean values and
\begin{align}
  \overline{ H_{\mu s_{\mu} \nu s_{\nu}}^{ij}
             H_{\mu' s_{\mu'} \nu' s_{\nu'}}^{ji} }
  & =   \frac{v_{2}^{2}}{2M}
         \delta_{\mu \nu'}\delta_{\nu \mu'}
         \delta_{s_{\mu} s_{\nu'}}\delta_{s_{\nu} s_{\mu'}} ,
            \\
  \overline{ H_{\mu s_{\mu} \nu s_{\nu}}^{ij}
             H_{\mu' s_{\mu'} \nu' s_{\nu'}}^{ij} }
  & = - \frac{v_{2}^{2}}{2M}
         \delta_{\mu \mu'}\delta_{\nu \nu'}
         ({\rm i}\sigma_{y})_{s_{\mu'},s_{\mu}}
         ({\rm i}\sigma_{y})_{s_{\nu},s_{\nu'}} .
\end{align}
All other matrix elements $H_{\mu s_{\mu} \nu s_{\nu}}^{ij}$ vanish.
The coupling to the leads is described by a fixed
$2KM \times (N_{1} + N_{2})$ matrix $W$.
Its elements $W_{\mu s_{\mu} a}^{i}$,
where $a$ denotes the modes in the leads,
are nonzero only for $i = 1$ and $1 \le a \le N_{1}$,
and $i = K$ and $N_{1}+1 \le a \le N_{1}+N_{2}$.
We assume that the orthogonality relation
$\pi \sum_{\mu s_{\mu}} W_{a \mu s_{\mu}}^{i}W_{\mu s_{\mu}b}^{i}
= v_{1} x \delta_{ab}$ holds for $i = 1, K$.~\cite{iida}
We decompose $W$ as $W=W_{1}+W_{2}$, where $W_{1}$ and $W_{2}$ contain
the block with $\{W_{\mu s_{\mu}a}^{1}\}$
and that with $\{W_{\mu s_{\mu}a}^{K}\}$, respectively.
Brouwer and Frahm proved that the generating function eq.~(\ref{eq:g-func1})
can be expressed in the form,
\begin{align}
        \label{eq:g-func2}
    F(\hat{\theta}, \hat{\lambda})
 & = {\rm Sdet}^{-\frac{1}{2}}
     \Big( E_{\rm F} 1_{8} - H 1_{8}
                  + {\rm i}\pi W_{1}W_{1}^{\dagger}Q
              \nonumber \\
 &  \hspace{30mm}
                  + {\rm i}\pi W_{2}W_{2}^{\dagger}\Lambda \Big) ,
\end{align}
where ${\rm Sdet}$ stands for the superdeterminant of a supermatrix
and $E_{\rm F}$ the Fermi energy.
The supermatrix $Q$ is expressed as $Q = T^{-1}\Lambda T$
with $\Lambda  = {\rm diag}(1_{4}, - 1_{4})$ and
\begin{equation}
    T  = \left( \begin{array}{cc}
                    u^{-1} & 0      \\
                    0      & v^{-1} \\
                \end{array} \right)
         \exp \left( \begin{array}{cc}
                         0 & \frac{1}{2} \hat{\theta} \\
                         \frac{1}{2} \hat{\theta} & 0 \\
                     \end{array} \right)
         \left( \begin{array}{cc}
                       u & 0 \\
                       0 & v \\
                \end{array} \right) ,
\end{equation}
where $u$ and $v$ are $4 \times 4$ unitary supermatrices and
\begin{equation}
  \hat{\theta}
   = \left( \begin{array}{cccc}
               \theta_{1} & 0 & 0 & 0 \\
               0 & \theta_{1} & 0 & 0 \\
               0 & 0 & {\rm i}\theta_{3} & {\rm i}\theta_{4} \\
               0 & 0 & {\rm i}\theta_{4} & {\rm i}\theta_{3} \\
            \end{array} \right) .
\end{equation}
We can rewrite $F$ in terms of a Gaussian integral over
supervector $\psi$ as
\begin{align}
    F(\hat{\theta}, \hat{\lambda})
 & = \int D \psi
     \exp \bigg[ \frac{\rm i}{2} \psi^{\dagger}\Lambda
          \Big( E_{\rm F} 1_{8} - H 1_{8}
                    + {\rm i}\pi W_{1}W_{1}^{\dagger}Q
                \nonumber \\
 & \hspace{27mm}    + {\rm i}\pi W_{2}W_{2}^{\dagger}\Lambda
                    + {\rm i}\epsilon \Lambda \Big) \psi \bigg] ,
\end{align}
where $\epsilon$ is a positive infinitesimal.
After performing the standard manipulations,~\cite{efetov,iida,mirlin}
we find that in the limit $M \to \infty$, 
the averaged generating function is given by
\begin{align}
   \langle F(\hat{\theta}, \hat{\lambda}) \rangle_{\rm SSF}
 & = \int {\rm d}Q_{1} \int {\rm d}Q_{K}f_{1}(Q,Q_{1})f_{2}(\Lambda,Q_{K})
          \nonumber \\
 & \hspace{20mm}  \times W(Q_{1},Q_{K}) ,
\end{align}
where
\begin{align}
     \label{eq:heat-k}
   W(Q_{1},Q_{K})
  & = \int \prod_{j=2}^{K-1}{\rm d}Q_{j}
             \nonumber \\
  & \hspace{-7mm}
      \times
      \exp \left( - (v_{2}/v_{1})^{2}
        \sum_{k=1}^{K-1}{\rm Str}\left(Q_{k}Q_{k+1} \right) \right) ,
         \\
   f_{1}(Q,Q_{1})
  & = \exp \left( - \frac{N_{1}}{2} {\rm Str} \ln (1+xQQ_{1}) \right) ,
        \\
   f_{2}(Q,Q_{K})
  & = \exp \left( - \frac{N_{2}}{2} {\rm Str} \ln (1+xQQ_{K}) \right) .
\end{align}
Here, ${\rm Str}$ denotes the supertrace for a supermatrix.
Assuming that $v_{1}^{2} \ll v_{2}^{2}$, we replace the sum
in eq.~(\ref{eq:heat-k}) with an integral.~\cite{iida,mirlin}
Accordingly, the discrete number $k$ is reduced to the continuous
variable $s$.
The propagator $W(Q,Q')$, called the heat kernel,
obeys the heat equation in the supersymmetric space.
Noting that $f_{1}(Q,Q')$ has the same symmetry as the heat kernel,
we eventually find that
$\langle F(\hat{\theta}, \hat{\lambda}) \rangle_{\rm SSF}$
 obeys~\cite{brouwer}
\begin{equation}
     \label{eq:F-smf}
  8(v_{2}/v_{1})^{2} \frac{\partial}{\partial s}
  \langle F(\hat{\theta}, \hat{\lambda}) \rangle_{\rm SSF}
    = \Delta_{\hat{\theta}}
      \langle F(\hat{\theta}, \hat{\lambda}) \rangle_{\rm SSF}
\end{equation}
with the initial condition
\begin{equation}
     \label{eq:init-smf1}
   \lim_{s \to 0} \langle F(\hat{\theta}, \hat{\lambda}) \rangle_{\rm SSF}
    = \int {\rm d}Q' f_{1}(Q,Q') f_{2}(\Lambda,Q') .
\end{equation}
The Laplacian is defined as
\begin{equation}
   \Delta_{\hat{\theta}}
  = \sum_{j=1,3,4} J^{-1}(\hat{\theta})\frac{\partial}{\partial \theta_{j}}
                   J(\hat{\theta})\frac{\partial}{\partial \theta_{j}}
\end{equation}
with
\begin{align}
   J(\hat{\theta})
 &  = \sin\theta_{3}\sin\theta_{4}\sinh^{3}\theta_{1}
          \nonumber \\
 &     \times \prod_{s_{1}, s_{2} = \pm 1}
       \sinh^{-2}\left[ \frac{1}{2}
           \left(\theta_{1}+{\rm i}s_{1}\theta_{3}+{\rm i}s_{2}\theta_{4}
           \right)\right] .
\end{align}

We prove that the SSF approach
is equivalent to the DMPK approach.
It is convenient to introduce $D_{\hat{\lambda}}^{(0)}$, which appears in
the evolution equation for the even-channel case,~\cite{brouwer} given by
\begin{equation}
    D_{\hat{\lambda}}^{(0)}
  = J_{0}^{-1}(\hat{\lambda}) \sum_{a=1}^{m}
    \frac{\partial}{\partial \lambda_{a}}
    \lambda_{a}(1+\lambda_{a}) J_{0}(\hat{\lambda})
    \frac{\partial}{\partial \lambda_{a}}
\end{equation}
with $J_{0}(\hat{\lambda}) =
\prod_{a=2}^{m} \prod_{b=1}^{a-1} |\lambda_{a}-\lambda_{b}|^{4}$.
Using the relation
$D_{\hat{\lambda}}^{(0)}F_{0}(\hat{\theta}, \hat{\lambda})
= \Delta_{\hat{\theta}}F_{0}(\hat{\theta}, \hat{\lambda})$,
derived by Brouwer and Frahm, we obtain
\begin{align}
   \Delta_{\hat{\theta}}F(\hat{\theta}, \hat{\lambda})    
 &  = \left( D_{\hat{\lambda}}^{(0)}
            + \sum_{a=1}^{m} \delta g(\hat{\theta}, \lambda_{a}) \right)
        F(\hat{\theta}, \hat{\lambda})
         \nonumber \\
 & \hspace{10mm}
    + \left( \Delta_{\hat{\theta}}
          f(\hat{\theta}, \lambda_{N})^{\frac{1}{2}}\right)
          F_{0}(\hat{\theta}, \hat{\lambda}) ,
\end{align}
where
\begin{equation}
   \delta g(\hat{\theta}, \lambda)
   = \sum_{j=1,3,4}
       2 \frac{\partial \ln f(\hat{\theta}, \lambda_{N})^{\frac{1}{2}}}
              {\partial \theta_{j}}
         \frac{\partial \ln f(\hat{\theta}, \lambda)}
              {\partial \theta_{j}} .
\end{equation}
We can show that
$\Delta_{\hat{\theta}}
f(\hat{\theta}, \lambda_{N})^{1/2}\big|_{\lambda_{N}=0} = 0$
and
\begin{equation}
  \left( D_{\hat{\lambda}}^{(0)}
          + \sum_{a=1}^{m} \delta g(\hat{\theta}, \lambda_{a}) \right)
  F(\hat{\theta}, \hat{\lambda})\big|_{\lambda_{N}=0}
  = D_{\hat{\lambda}}F(\hat{\theta}, \hat{\lambda})\big|_{\lambda_{N}=0} ,
\end{equation}
which result in 
\begin{equation}
   \Delta_{\hat{\theta}} F(\hat{\theta}, \hat{\lambda})\big|_{\lambda_{N}=0}
     = D_{\hat{\lambda}}F(\hat{\theta}, \hat{\lambda})\big|_{\lambda_{N}=0} .
\end{equation}
Using this relation, we obtain
\begin{align}
 & l(N-1) \frac{\partial}{\partial L}
   \langle F(\hat{\theta}, \hat{\lambda}) \rangle_{\rm DMPK}
                    \nonumber \\
 & \hspace{5mm}
   = \int_{0}^{\infty} \prod_{a=1}^{m}{\rm d}\lambda_{a}
      \int_{0}^{\infty} {\rm d}\lambda_{N} \delta(\lambda_{N})
      P(\hat{\lambda},L)
      D_{\hat{\lambda}} F(\hat{\theta}, \hat{\lambda})
                    \nonumber \\
 & \hspace{5mm}
   = \int_{0}^{\infty} \prod_{a=1}^{m}{\rm d}\lambda_{a}
      \int_{0}^{\infty} {\rm d}\lambda_{N}
      \delta(\lambda_{N})P(\hat{\lambda},L)
      \Delta_{\hat{\theta}} F(\hat{\theta}, \hat{\lambda})
                    \nonumber \\
 & \hspace{5mm}
   = \Delta_{\hat{\theta}}
      \langle  F(\hat{\theta}, \hat{\lambda}) \rangle_{\rm DMPK} .
\end{align}
Thus, the evolution equation for
$\langle  F(\hat{\theta}, \hat{\lambda}) \rangle_{\rm DMPK}$
is equivalent to that for
$\langle  F(\hat{\theta}, \hat{\lambda}) \rangle_{\rm SSF}$ if we set
\begin{equation}
  \frac{8}{s}\left(\frac{v_{2}}{v_{1}}\right)^{2}
    = \frac{l(N-1)}{L} \equiv \frac{\xi}{2L} .
\end{equation}

Now we focus on the initial condition.
We assume that the coupling to the two leads is ideal,
which is achieved by setting $x = 1$.
Brouwer and Frahm showed that the ballistic initial condition
for the DMPK approach (i.e., $\lambda_{a}=0$ for any $a$) can be reproduced
if we fix $N_{1}$ and take the limit $N_{2} \to \infty$.
We trace the same steps.
The limit $N_{2} \to \infty$ results in
$f_{2}(\Lambda,Q') \to \delta(Q',\Lambda)$.
Thus, the right-hand side of eq.~(\ref{eq:init-smf1})
is reduced to $f_{1}(Q,\Lambda)$.
We finally obtain
\begin{align}
        \label{eq:init-smf2}
   \lim_{s \to 0} \langle F(\hat{\theta}, \hat{\lambda}) \rangle_{\rm SSF}
  & = \exp \left( - \frac{N_{1}}{2} {\rm Str} \ln (1+Q\Lambda) \right)
           \nonumber \\
  & = \left( \frac{\cos \theta_{3}+\cos \theta_{4}}
                {1+\cosh \theta_{1}} \right)^{N_{1}} .
\end{align}
This is equivalent to eq.~(\ref{eq:init-dmpk}) if we set
$N_{1}= N = 2m + 1$.
Thus, the SSF approach is equivalent to the DMPK approach
if we accept the treatments described above.
This indicates that the average conductance as well as its variance is
the same whether computed from the SSF approach or from the DMPK approach.
It should be emphasized that eq.~(\ref{eq:init-smf2}) becomes
the initial condition for the even-channel case if we set $N_{1} = 2m$.

We here briefly summarize the Zirnbauer's super-Fourier
analysis~\cite{zirnbauer,mirlin} for the thick-wire limit which is defined
by $N \to \infty$, $L/l \to \infty$ with a fixed $Nl/L$.
In this analysis, we expand the heat kernel $W(Q,Q')$
in terms of the eigenfunctions of $\Delta_{\hat{\theta}}$.
The eigenvalue equation is~\cite{zirnbauer,mirlin,brouwer}
\begin{equation}
  \Delta_{\hat{\theta}}\phi_{\lambda,n_{1},n_{2}}(\hat{\theta})
    = - \frac{1}{4}\varepsilon (\lambda,n_{1},n_{2})
        \phi_{\lambda,n_{1},n_{2}}(\hat{\theta}) .
\end{equation}
The eigenvalues are
$\varepsilon (\lambda,n_{1},n_{2}) = \lambda^{2}+n_{1}^{2}+n_{2}^{2}-1$,
where $n_{1},n_{2} = 1,3,5,\cdots$ and $\lambda$ is a positive real number.
In addition, there arises the zero mode $\varepsilon ({\rm i},1,1) = 0$,
and the subsidiary series
$\varepsilon ({\rm i},n,n-2) = 2(n-1)^{2}$, where $n=3,5,7,\cdots$.
Using the resulting expansion, Zirnbauer obtained the expression
of the average conductance,
\begin{align}
  \langle g \rangle_{\rm Zirnbauer}
 & =  T({\rm i},1,1)/2
             \nonumber \\
 & \hspace{-10mm}
    + \sum_{n=3,5,7,\cdots}
      \left( T({\rm i},n,n-2) + T({\rm i},n-2,n) \right)/2
           \nonumber \\
 & \hspace{-10mm}
    + 2^{4} \sum_{n_{1},n_{2}=1,3,5,\cdots}
      \int_{0}^{\infty}{\rm d}\lambda \lambda (\lambda^{2}+1)
      \tanh(\pi \lambda/2)
           \nonumber \\
 & \hspace{-3mm}
      \times n_{1}n_{2}(\lambda^{2}+n_{1}^{2}+n_{2}^{2}-1)
           \nonumber \\
 &  \hspace{-3mm} \times
      \prod_{\sigma,\sigma_{1},\sigma_{2}=\pm 1}
      (- 1 + {\rm i}\sigma \lambda + \sigma_{1}n_{1} + \sigma_{2}n_{2})^{-1}
           \nonumber \\
 &  \hspace{-3mm} 
      \times T(\lambda,n_{1},n_{2})
\end{align}
with
$T(\lambda,n_{1},n_{2}) =
{\rm e}^{-(\lambda^{2}+n_{1}^{2}+n_{2}^{2}-1) (L/2\xi)}$.
This indicates that $\langle g \rangle \to 1/2$
when $L \to \infty$.~\cite{zirnbauer,mirlin}
This apparently contradicts the DMPK approach, which gives
$\langle g \rangle \to 0$ in the even-channel case~\cite{beenakker}
and $\langle g \rangle \to 1$ in the odd-channel case.~\cite{takane1}

Note that the difference between the even- and odd-channel cases
is not reflected in the Zirnbauer's argument.
We show that the even-odd difference can be taken into account
through the initial condition for the heat kernel.
Let us consider an operation $\hat{P}$ defined by
\begin{equation}
   \hat{P}\phi(\theta_{1},\theta_{3},\theta_{4})
    = \phi(\theta_{1},\theta_{3}+\pi,\theta_{4}+\pi) ,
\end{equation}
which is introduced by Brouwer and Frahm.~\cite{brouwer}
The Laplacian $\Delta_{\hat{\theta}}$ commutes with $\hat{P}$,
and the eigenfunctions have either even or odd parity.
The even (odd) parity means that
the eigenvalue of $\hat{P}$ is $+1$ ($-1$).
Zirnbauer considered the limit $N_{1} = N_{2} \to \infty$.
If the same limit is taken in our argument, we find
that $\langle F \rangle_{\rm SSF} = W(Q,\Lambda)$ holds~\cite{brouwer}
since $f_{i}(Q,Q')$ is reduced to the delta function $\delta (Q,Q')$
when $N_{i} \to \infty$.
This indicates that we can obtain the heat kernel by
expanding $\langle F \rangle_{\rm SSF}$ in terms of the eigenfunctions
of $\Delta_{\hat{\theta}}$
and then taking the limit $N_{1} \to \infty$.
By employing this method, we can obtain
$W(Q,\Lambda)$ which reflects the even-odd difference.
The crucial point is that before taking the limit $N_{1} \to \infty$,
the parity of eq.~(\ref{eq:init-smf2}) is well defined;
its parity is odd (even) when $N_{1} = 2m + 1$ ($N_{1} = 2m$).
Accordingly, we must select only the eigenmodes having odd parity
in the odd-channel case, while the even-parity modes are appropriate
in the even-channel case.
However, once the limit of $N_{1} \to \infty$ is taken,
eq.~(\ref{eq:init-smf2}) is reduced to
\begin{equation}
     \label{eq:init-after}
   \lim_{s \to 0} \langle F \rangle_{\rm SSF} = \delta (Q,\Lambda),
\end{equation}
which is equivalent to the Zirnbauer's initial condition,
and we cannot derive such a parity-selection rule from it.
Thus, the even-odd difference disappears.
In order to correctly calculate the conductance,
we must take care of this parity-selection rule.
Brouwer and Frahm pointed out that the odd-parity modes are
unphysical in the even-channel case.
However, they were not aware that the odd-parity modes become essential
in the odd-channel case.

In the even-channel case, the above argument tells us that
we must neglect the eigenfunctions with odd parity.
The zero mode as well as its subsidiary modes has
odd parity,~\cite{brouwer} so they must be excluded.
By omitting all the odd-parity modes and correctly taking account of
Kramers degeneracy, Brouwer and Frahm obtained
\begin{equation}
      \label{eq:av-g-even}
   \langle g \rangle \sim \frac{32}{9}
   \left(\frac{\pi \xi}{2L}\right)^{\frac{3}{2}}
   {\rm e}^{-\frac{L}{2 \xi}} ,
\end{equation}
which agrees with the DMPK approach.~\cite{beenakker}
In contrast, we must exclude the even-parity modes in the odd-channel case.
This is the crucial difference between the ordinary even-channel case
and our odd-channel case.
If we omit all the even-parity modes and take Kramers degeneracy
into account, we obtain
\begin{align}
  \langle g \rangle
 & =  T({\rm i},1,1)
    + \sum_{n=3,5,7,\cdots}\left( T({\rm i},n,n-2) + T({\rm i},n-2,n) \right)
           \nonumber \\
 & \hspace{5mm}
    + 2^{5} \sum_{\scriptstyle n_{1},n_{2}=1,3,5,\cdots
                  \atop \scriptstyle n_{1}+n_{2}\equiv 0 ({\rm mod \, 4})}
      \int_{0}^{\infty}{\rm d}\lambda \lambda (\lambda^{2}+1)
      \tanh(\pi \lambda/2)
            \nonumber \\
 & \hspace{15mm}
      \times n_{1}n_{2}(\lambda^{2}+n_{1}^{2}+n_{2}^{2}-1)
           \nonumber \\
 &  \hspace{15mm}
      \times \prod_{\sigma,\sigma_{1},\sigma_{2}=\pm 1}
      (- 1 + {\rm i}\sigma \lambda + \sigma_{1}n_{1} + \sigma_{2}n_{2})^{-1}
           \nonumber \\
 &  \hspace{15mm}
      \times T(\lambda,n_{1},n_{2}) .
\end{align}
When $L \to \infty$, the leading contribution to $\langle g \rangle$ arises
from the zero mode, which yields $\langle g \rangle \to 1$.
This indicates the presence of one perfectly conducting channel.
The deviation $\langle \delta g \rangle \equiv \langle g \rangle - 1$
from the quantized value is
\begin{equation}
   \langle \delta g \rangle \sim 2 {\rm e}^{-\frac{4 L}{\xi}} ,
\end{equation}
which comes from the subsidiary modes with $n=3$.
This clearly indicates that the zero mode as well as
the subsidiary modes is essential in describing
the anomalous transport property in the odd-channel case.
The characteristic length scale for $\langle \delta g \rangle$
is $\xi_{\rm odd} = \xi /4$, while that in the even-channel case
is obtained as $\xi_{\rm even} = 2 \xi$ from eq.~(\ref{eq:av-g-even}).
Thus, $\xi_{\rm odd}$ is a factor $8$ shorter than $\xi_{\rm even}$.
This result is in quantitative agreement
with the DMPK approach.~\cite{takane1}


\begin{thebibliography}{99}
\bibitem{comment} Here we have assumed that $N$ represnts
the number of conducting channels including Kramers degeneracy.
That is, $g = 2 \sum_{a=1}^{m}T_{a}$ with $m = N/2$.
\end{thebibliography}
\end{document}